\documentclass[pra,twocolumn,showpacs,preprintnumbers,amsmath,amssymb]{revtex4}
\usepackage{mathrsfs}
\usepackage{bbm}
\usepackage{amsfonts}
\usepackage{tipa}

\usepackage{epsfig,graphicx}
\usepackage{amstext}
\usepackage{amsmath}
\usepackage{graphicx}
\usepackage{times}

\begin{document}


\title{Measurement-induced nonlocality based on the trace norm}

\author{Ming-Liang Hu$^{1}$}
\email{mingliang0301@163.com}
\author{Heng Fan$^{2}$}
\email{hfan@iphy.ac.cn}
\affiliation{$^{1}$School of Science, Xi'an University of Posts and
               Telecommunications, Xi'an 710121, China \\
             $^{2}$Beijing National Laboratory for Condensed Matter Physics,
               Institute of Physics, Chinese Academy of Sciences, Beijing
               100190, China}

\begin{abstract}
Nonlocality is one unique property of quantum mechanics differing
from classical world. One of its quantifications can be properly
described as the maximum global effect caused by locally invariant
measurements, termed as measurement-induced nonlocality (MIN) (2011
\emph{Phys. Rev. Lett.} {\bf 106} 120401). Here, we propose to
quantify the MIN by the trace norm. We show explicitly that this
measure is monotonically decreasing under the action of completely
positive trace-preserving map, which is the general local quantum
operation, on the unmeasured party for the bipartite state. This
property avoids the undesirable characteristic appearing in the
known measure of MIN defined by the Hilbert-Schmidt norm that may be
increased or decreased by trivial local reversible operations on the
unmeasured party. We obtain analytical formulas of the trace-norm
MIN for any $2\times n$ dimensional pure state, two-qubit state, and
certain high-dimensional states. As other quantum correlation
measures, the new defined MIN can be directly applied to various
models for physical interpretations.

\end{abstract}

\pacs{03.65.Ud, 03.65.Ta, 03.67.Mn
      \\Key Words: Measurement-induced nonlocality; Trace norm;
                   Quantum correlations
     }

\maketitle

\section{Introduction}\label{sec:1}
Quantum physics differs in many aspects from our conventional
intuition. One of such intriguing difference is the celebrated
notion of nonlocality, which arises from the debate of the early
20th century among scientists. One line of the debate originated
with Einstein, Podolsky, and Rosen, who proposed the thought
experiment known as EPR paradox and the so-called
``spooky-action-at-a-distance" \cite{locality}. Their predictions of
quantum mechanics are, in sharp contrast to the conventional view,
that physical processes should obey the principle of locality.

Nonlocality of quantum physics has been studied from different
points of view, such as by Bell inequality and entanglement. The
Bell-type inequalities \cite{Bell,chsh,Augusiak} are derived by the
local hidden variable theory, and may be violated by the quantum
measurement outcomes realizable experimentally \cite{physrep}. The
violation of Bell inequalities implies the existence of entanglement
in a system \cite{chsh}, however, this is not always true for the
opposite case, as there exist mixed states that are entangled but do
not violate any Bell-type inequalities \cite{Werner}. Other studies
further consolidate that there is also nonlocality without
entanglement \cite{nowen}, or nonlocality without quantum
correlations other than entanglement \cite{nowqu}. So nonlocality
seems can be quantified reasonably by different measures.

Nowadays, we realized that nonlocality is not only a central concept
of quantum mechanics, but may also be used to improve the efficiency
of many quantum information processing (QIP) tasks \cite{RMP,qip}.
Meanwhile, it is also interrelated with other foundational theory of
quantum mechanics such as uncertainty principle
\cite{Oppenheim-science}. These delicate and intriguing features of
nonlocality prompted a huge surge of people's interest from the
quantum physics community, with notable progresses being achieved in
the past few years \cite{glo1,glo2,glo3,new1}.

Apart from the traditional line for quantum nonlocality related to
entanglement or Bell inequalities, it is also significant to
investigate it from other perspectives. Recently, Luo and Fu
\cite{min} presented a new measure of nonlocality which they termed
as measurement-induced nonlocality (MIN) motivated by the definition
of quantum discord \cite{Ollivier,new2}. As the name itself
indicates, the MIN characterizes nonlocality from a measurement
perspective, and thus is different from other nonlocality measures.
It is a manifestation of the global disturbance to the overall state
of a system caused by the locally non-disturbing measurement on one
subsystem, and can also be considered as one kind of nonclassical
correlation measure different from entanglement and quantum discord.

\section{MIN quantified by Hilbert-Schmidt norm}\label{sec:2}
The MIN has been one of current research focuses for years
\cite{min2,min3,min4,min5, min6,min7,min8,min9,uin}. However, its
quantification based on the Hilbert-Schmidt norm (we call it
conventional MIN for brevity), while intuitively appealing and
conceptually significant, has certain discouraging properties. To
see this explicitly, we recall its definition, which reads
\cite{min}
\begin{equation}\label{eq1}
 N_2(\rho_{AB})=\max_{\Pi^A}||\rho_{AB}-\Pi^A(\rho_{AB})||_2^2,
\end{equation}
for a bipartite state $\rho_{AB}$ in $\mathcal {H}_{AB}$. Here,
$||X||_2=\sqrt{\text{Tr}(X^\dag X)}$ denotes the Hilbert-Schmidt
norm, and the maximum is taken over the full set of local projective
measurements $\Pi^A=\{\Pi_k^A\}$ that keep the reduced state
$\rho_A=\text{Tr}_B \rho_{AB}$ invariant, namely, $\sum_k
\Pi_k^A\rho_A\Pi_k^A=\rho_A$. An analytical formula of the
conventional MIN for any $2\times n$ dimensional state $\rho_{AB}$
can be obtained \cite{min}. Here, we argue that the conventional MIN
in Eq. \eqref{eq1}, despite being favored for its convenience of
calculation, may has certain undesirable properties. More
specifically, we will show that it can increase or decrease under
trivial local reversible operations on the unmeasured subsystem $B$
of $\rho_{AB}$. Consider, for instance, a channel $\Gamma_B$ acting
as $\Gamma_B(\rho_{AB}) = \rho_{AB}\otimes\rho_C$ (i.e., it
introduces a local ancilla to $B$), then by making use of the
multiplicativity of the Schatten $p$-norm (which reduces to the
Hilbert-Schmidt norm when $p=2$) under tensor products, we obtain
\begin{equation}\label{eq2}
 N_2(\rho_{A:BC})=N_2(\rho_{AB})\text{Tr}\rho_C^2.
\end{equation}
This equality means that $N_2(\rho_{A:BC})\le N_2(\rho_{AB})$ as the
purity of a state is no larger than one, $\text{Tr}\rho_C^2 \le 1$.
Particularly, if $\rho_C = \mathbb{I}_n/n$ with $\mathbb{I}_n$ being
the $n$-dimensional identity operator, we obtain $N_2(\rho_{A:BC})=
N_2(\rho_{AB})/n$. Then $N_2(\rho_{A:BC})$ will approach zero when
$n$ takes the limit of infinity. This behavior differs completely
from our intuition that the nonlocal properties of a system should
not be affected by trivially adding or removing an uncorrelated
local ancillary state.

We remark here that the above perplexity is reminiscent of the
phenomena encountered for the geometric measure of quantum discord
(GQD) \cite{gqd,gqd1,Piani,Dakic}. In that case, several
well-defined measures of GQD have been introduced to remedy this
problem \cite{lqu,square,trace,Ciccarello,bures}.

\section{MIN based on the trace norm}\label{sec:3}
Motivated by the proposition for modifying GQD via trace norm
\cite{trace}, we propose to define the MIN for a bipartite state
$\rho_{AB}$ as
\begin{equation}\label{eq3}
 N_1(\rho_{AB})=\max_{\Pi^A}||\rho_{AB}-\Pi^A(\rho_{AB})||_1,
\end{equation}
where $||X||_1= \text{Tr}\sqrt{X^\dag X}$, and $\Pi^A$ denotes still
the projective measurements that satisfy $\Pi^A(\rho_A)=\rho_A$. We
call $N_1(\rho_{AB})$ the trace MIN hereafter. The physical
interpretation of this new nonlocality measure can still be
presented as the maximal global effect, or more explicitly, the
maximal trace distance that the postmeasurement state
$\Pi^A(\rho_{AB})$ departs from its premeasurement state
$\rho_{AB}$, caused by locally invariant measurements.

The nonlocality measure defined above can circumvent the problem
occurred for the conventional MIN as implied by Eq. \eqref{eq2}.
Explicitly, let us repeat the analysis by adding an uncorrelated
ancilla when the new definition is used, then $N_1(\rho_{A:BC})
=N_1(\rho_{AB})$ due to the normalization condition $\text{Tr}\rho_C
=1$. Therefore, $N_1(\rho_{AB})$ does not increase under the action
of $\Gamma_B$, namely, it is unaffected by adding or removing a
factorized local ancilla on the unmeasured party. Here, we further
show a more general and powerful result related to the trace MIN in
Eq. \eqref{eq3}.

\emph{Theorem 1.} The trace MIN $N_1(\rho_{AB})$ defined in Eq.
\eqref{eq3} is nonincreasing under the action of any completely
positive trace-preserving (CPTP) channel $\mathcal {E}_B$ on the
unmeasured party $B$, i.e., we always have
\begin{equation}\label{eq4}
 N_1(\rho_{AB})\geqslant N_1[\mathcal {E}_B(\rho_{AB})].
\end{equation}

\emph{Proof.} Let $\mathcal {E}_B$ be an arbitrary CPTP channel
acting on party $B$ of $\rho_{AB}$, and $\{\tilde{\Pi}_k^A\}$ be the
optimal projection-valued measurement on party $A$ that maximizes
the trace norm on the right-hand side of Eq. \eqref{eq3} for
$N_1[\mathcal {E}_B(\rho_{AB})]$, namely,
\begin{equation}\label{eq-theo11}
 N_1[\mathcal{E}_B(\rho_{AB})] = ||\mathcal {E}_B(\rho_{AB})
                                 -\tilde{\Pi}^A[\mathcal{E}_B(\rho_{AB})]||_1,
\end{equation}
then, by noting that any local channel on party $B$ and the
measurement made on party $A$ commute, we obtain $\tilde{\Pi}^A
[\mathcal{E}_B (\rho_{AB})]=\mathcal{E}_B [\tilde{\Pi}^A
(\rho_{AB})]$, and therefore, by denoting $\bar{\Pi}^A$ (note that
$\bar{\Pi}^A \neq \tilde{\Pi}^A$ in general) the optimal measurement
for obtaining $N_1(\rho_{AB})$, we have
\begin{eqnarray}\label{eq-theo12}
 N_1(\rho_{AB})&=&||\rho_{AB}-\bar{\Pi}^A (\rho_{AB})||_1 \nonumber\\
               &\geqslant& ||\rho_{AB}-\tilde{\Pi}^A (\rho_{AB})||_1 \nonumber\\
               &\geqslant& ||\mathcal{E}_B(\rho_{AB})
                -\mathcal{E}_B [\tilde{\Pi}^A (\rho_{AB})]||_1 \nonumber\\
               &=&N_1[\mathcal {E}_B(\rho_{AB})],
\end{eqnarray}
where the first inequality comes from the fact that $\tilde{\Pi}^A
(\rho_{AB})$ is not necessarily the optimal state to $\rho_{AB}$,
and the second inequality is due to the contractivity of the trace
norm under CPTP map (Theorem 9.2 of Ref. \cite{Nielsen}). This
completes the proof. \hfill{$\blacksquare $}

The above theorem means that no physical process on $B$ can increase
the maximum trace distance between a state $\rho_{AB}$ and its
postmeasurement state $\Pi^A(\rho_{AB})$ obtained after the locally
invariant measurements on party $A$, and therefore it circumvents
successfully the problem occurred for the conventional MIN.

We now list some other basic properties of the trace MIN. (i)
$N_1(\rho_{AB})=0$ for all the product states $\rho_{AB}=\rho_A
\otimes \rho_B$, and the classical-quantum states $\rho_{AB} =\sum_k
p_k \Pi_k^A\otimes \rho^B_k$ with nondegenerate $\rho_A= \sum_k p_k
\Pi_k^A$. (ii) $N_1(\rho_{AB})$ is invariant under locally unitary
operation $U=V_A\otimes W_B$ on $\rho_{AB}$, namely, $N_1
(U\rho_{AB}U^\dagger)=N_1(\rho_{AB})$, which is obvious as the trace
norm is preserved under unitary transformations \cite{Nielsen}.

The proposed trace MIN can be used to detect the effect of a locally
invariant measurement on the overall state of a system, and the zero
trace MIN implies that the state of the system cannot be disturbed
by any locally invariant measurement, namely, the measurement of one
subsystem cannot determine the corresponding result of a measurement
of the other, and therefore this system obeys the principle of
locality. Moreover, from the basic properties listed above, one can
note that while any entangled or discordant state possess
nonvanishing trace MIN, there also exist states with nonvanishing
trace MIN but do not produce correlations of entanglement or
discord. Therefore, the trace MIN is an important complementary to,
but different from, entanglement and quantum discord.

\section{Analytical formulas of the trace MIN}\label{sec:4}
The maximization in Eq. \eqref{eq3} over the full set of locally
invariant measurements on party $A$ can be obtained for certain
family of states, and in turn the trace MIN can be evaluated
analytically. We present them via the following theorems.

\emph{Theorem 2.} For any $2\times n$ dimensional pure state
$|\psi\rangle$ with the Schmidt decomposition
$|\psi\rangle=\sum_{k=1}^2 \sqrt{\lambda_k}|\phi^A_k\rangle\otimes
|\phi^B_k\rangle$, the trace MIN is given by
\begin{eqnarray}\label{eq6}
 N_1(|\psi\rangle\langle\psi|)=2\sqrt{\lambda_1\lambda_2}.
\end{eqnarray}

\emph{Proof.} We denote $\rho^\psi=|\psi\rangle\langle\psi|$, and
$\rho_A^\psi= \text{Tr}_B \rho^\psi$ for simplicity, then if
$\rho_A^\psi$ is nondegenerate, the optimal measurement
$\tilde{\Pi}_k^A =|\phi^A_k\rangle\langle \phi^A_k|$, and we have
\begin{eqnarray}\label{eq-theo21}
 \tilde{\Pi}^A (\rho^\psi)=\sum_{k=1}^2 \lambda_k |\phi^A_k\rangle\langle
                   \phi^A_k|\otimes|\phi^B_k\rangle\langle\phi^B_k|,
\end{eqnarray}
therefore
\begin{eqnarray}\label{eq-theo22}
 \rho^\psi-\tilde{\Pi}^A (\rho^\psi)&=&\sqrt{\lambda_1\lambda_2}(|\phi^A_1\rangle\langle\phi^A_2|
                                       \otimes |\phi^B_1\rangle\langle\phi^B_2|\nonumber\\
                                    && +|\phi^A_2\rangle\langle \phi^A_1|
                                       \otimes |\phi^B_2\rangle \langle \phi^B_1|),
\end{eqnarray}
the singular values of which can be obtained as $\epsilon_{1,2} =
\sqrt{\lambda_1\lambda_2}$, and thus
\begin{equation}\label{eq-theo23}
 N_1(\rho^\psi)= 2\sqrt{\lambda_1\lambda_2}.
\end{equation}

If $\rho_A^\psi$ is degenerate (i.e., $\lambda_{1,2}=1/2$), assuming
the optimal locally invariant measurement to be
$\tilde{\Pi}_k^A=|\tilde{k}\rangle \langle \tilde{k}|$, with
\begin{equation}\label{eq-theo24}
 |\tilde{k}\rangle=a_1^k |\phi^A_1\rangle+a_2^k |\phi^A_2\rangle,
\end{equation}
then as one can always find a unitary operator $U_A$ such that $U_A
|\phi^A_k\rangle=|\tilde{k}\rangle$, and as $N_1(\rho^\psi)$ is
locally unitary invariant, we obtain $N_1(\rho^\psi)= 1 $ after a
similar analysis as that performed for the nondegenerate case, and
this completes our proof. \hfill{$\blacksquare$}

As the entanglement of formation (EoF) for $|\psi\rangle$ was given
by $E_f=-\sum_{k=1}^2 \lambda_k\log_2\lambda_k$ \cite{EoF}, the
above theorem implies that $N_1(|\psi\rangle\langle\psi|)$
constitutes an entanglement monotone. But this is not the case for
general states. Moreover, one can derive
$N_1(|\psi\rangle\langle\psi|)=\sqrt{2 N_2
(|\psi\rangle\langle\psi|)}$, which means that for this special
case, both $N_1(|\psi\rangle\langle\psi|)$ and
$N_2(|\psi\rangle\langle\psi|)$ give qualitatively the same
characterizations of nonlocality.

For general $m \times n$ dimensional pure state in the Schmidt
expression $|\Psi\rangle =\sum_{k=1}^{d} \sqrt{\lambda_k}
|\phi^A_k\rangle\otimes |\phi^B_k\rangle$ with $d=\min\{m,n\}$, and
$\rho^\Psi = |\Psi\rangle\langle\Psi|$, we have
\begin{eqnarray}\label{eq-theo25}
 \rho^\Psi-\tilde{\Pi}^A (\rho^\Psi)= \sum_{i\neq j}\sqrt{\lambda_i\lambda_j}
                              |\phi^A_i\rangle\langle \phi^A_j|
                              \otimes |\phi^B_i\rangle\langle \phi^B_j|,
\end{eqnarray}
when $\rho_A^\Psi= \text{Tr}_B \rho^\Psi$ is nondegenerate, but a
closed form of its singular values cannot be derived for
$m\geqslant3$, and in turn it is difficult to obtain an analytical
formula of $N_1(\rho^\Psi)$. For the degenerate $\rho_A^\Psi$, an
analysis similar as that for $m=2$ yields $N_1(\rho^\Psi)=
2(m-1)/m$.

Now, we calculate the trace MIN for a general two-qubit state
$\tau_{AB}$, which has been proved to be locally unitary equivalent
to $\rho_{AB}$ of the following form \cite{Horodecki}
\begin{equation}\label{eq7}
 \rho_{AB}=\frac{1}{4}\bigg(\mathbb{I}_2\otimes\mathbb{I}_2+
           \vec{x}\cdot\vec{\sigma}\otimes\mathbb{I}_2+
           \mathbb{I}_2\otimes\vec{y}\cdot\vec{\sigma}+
           \sum_{i=1}^3 c_i\sigma_i\otimes\sigma_i\bigg),
\end{equation}
where the vectors $\vec{x}=(x_1,x_2,x_3)$, $\vec{y} =(y_1,y_2,y_3)$,
and $x_{i}= \text{Tr}\rho_{AB}(\sigma_i\otimes \mathbb{I}_2)$,
$y_{i}= \text{Tr}\rho_{AB}(\mathbb{I}_2 \otimes \sigma_i)$,
$c_{i}=\text{Tr}\rho_{AB}(\sigma_i\otimes\sigma_i)$, and
$\sigma_{1,2,3}$ are the three Pauli operators.

The local unitary invariance of the trace MIN enables
$N_1(\tau_{AB})= N_1(\rho_{AB})$, and therefore it suffices to
consider the representative family of states $\rho_{AB}$ in Eq.
\eqref{eq7}, for which $N_1(\rho_{AB})$ can be evaluated
analytically.

\emph{Theorem 3.} For any two-qubit state of the form of Eq.
\eqref{eq7}, the trace MIN can be obtained as
\begin{equation}\label{eq8}
  N_1(\rho_{AB})=\left\{
  \begin{aligned}
   &\frac{\sqrt{\chi_{+}}+\sqrt{\chi_{-}}}{2||\vec{x}||_1}
                              &\text{if}~\vec{x}\neq 0,\\
   &\max\{|c_1|,|c_2|,|c_3|\} &\text{if}~\vec{x}=0,
  \end{aligned} \right.
\end{equation}
where $\chi_\pm=\alpha\pm 2\sqrt{\beta} ||\vec{x}||_1$, with
$\alpha=||\vec{c}||_1^2||\vec{x}||_1^2-\sum_i c_i^2x_i^2$,
$\vec{c}=(c_1,c_2,c_3)$, $\beta=\sum_{\langle
ijk\rangle}x_i^2c_j^2c_k^2$, and the summation runs over all the
cyclic permutations of $\{1,2,3\}$.

\emph{Proof.} If $\rho_A=(\mathbb{I}_2 +\vec{x}\cdot \vec{\sigma})/2$
is nondegenerate, that is, if $\vec{x}\neq 0$, then
the unique projective measurement leaving $\rho_A$ invariant is
induced by its spectral resolutions
\begin{eqnarray}\label{eq-theo21}
 \tilde{\Pi}_{1,2}^A=\frac{1}{2} \left(\mathbb{I}_2 \pm \frac{\vec{x}\cdot
             \vec{\sigma}}{||\vec{x}||_1}\right).
\end{eqnarray}
Thus we have \cite{glo3}
\begin{eqnarray}\label{eq-theo22}
 \tilde{\Pi}^A(\rho_{AB})&=&\frac{1}{4}\bigg(\mathbb{I}_2\otimes\mathbb{I}_2+
                    \vec{x}\cdot\vec{\sigma}\otimes\mathbb{I}_2+
                    \mathbb{I}_2\otimes\vec{y}\cdot\vec{\sigma}\nonumber\\
                &&+\frac{\vec{x}\cdot \vec{\sigma}}{||\vec{x}||_1^2}
                    \otimes \sum_{i=1}^3 c_i x_i \sigma_i\bigg),
\end{eqnarray}
and therefore
\begin{eqnarray}\label{eq-theo23}
 \rho_{AB}-\tilde{\Pi}^A(\rho_{AB})&=&\frac{1}{4}\bigg(\sum_{i=1}^3 c_i\sigma_i
                              \otimes\sigma_i\nonumber\\
                           && -\frac{\vec{x}\cdot \vec{\sigma}}{||\vec{x}||_1^2}
                              \otimes \sum_{i=1}^3 c_i x_i \sigma_i\bigg).
\end{eqnarray}

After a straightforward algebra, one can obtain the singular values
of $\rho_{AB}-\tilde{\Pi}^A(\rho_{AB})$ as
\begin{equation}\label{eq-theo24}
 \varepsilon_{1,2}=\frac{\sqrt{\chi_{+}}}{4||x||_1},~~
 \varepsilon_{3,4}=\frac{\sqrt{\chi_{-}}}{4||x||_1},
\end{equation}
with $\chi_{\pm}$ being given below Eq. \eqref{eq8}, and therefore
\begin{equation}\label{eq-theo25}
 N_1(\rho_{AB})=\frac{\sqrt{\chi_{+}}+\sqrt{\chi_{-}}}{2||\vec{x}||_1}.
\end{equation}

If $\rho_A$ is degenerate, i.e., $\vec{x}=0$, then by adopting the
similar lines as in Ref. \cite{Ciccarello}, one can obtain
\begin{equation}\label{eq-theo26}
 N_1(\rho_{AB})=\frac{1}{2}\sqrt{2\max_{\hat{e}}h(\hat{e})},
\end{equation}
with $\hat{e}=(\sin\theta\cos\phi,\sin\theta\sin\phi,\cos\theta)$
being a unit vector in $\mathbb{R}^3$. By further ordering the
singular values of the correlation tensor
$\mathcal{R}=\text{diag}\{c_1,c_2,c_3\}$ as $c_+ \geqslant c_0
\geqslant c_-$, we have \cite{Ciccarello}
\begin{equation}\label{eq-theo27}
 h(\hat{e})=Q+\sqrt{H},
\end{equation}
where $Q$ and $H$ are given by
\begin{eqnarray}\label{eq-theo28}
 &&Q=c_+^2+c_0^2-\sin^2\theta\left[c_0^2-c_-^2+
      \cos^2\phi(c_+^2-c_0^2)\right],\nonumber\\
 &&H=A(\theta)\sin^4\phi+B(\theta)\sin^2\phi+C(\theta),
\end{eqnarray}\\
with the $\theta$-dependent functions $A(\theta)$, $B(\theta)$, and
$C(\theta)$ (note that there is a misprint in \cite{Ciccarello},
$\gamma_2^2+\gamma_3^2$ in the expression for $C(\theta)$ should be
$\gamma_3^2-\gamma_2^2$) being given by
\begin{eqnarray}\label{eq-theo29}
 &&A(\theta)=\sin^4\theta(c_+^2-c_0^2)^2, \nonumber\\
 &&B(\theta)=2(c_+^2-c_0^2)[\sin^2\theta(c_+^2+c_0^2-2c_-^2) \nonumber\\
             &&~~~~~~~~~~~~~-\sin^4\theta(c_+^2-c_-^2)],\nonumber\\
 &&C(\theta)=[c_+^2-c_0^2-\sin^2\theta(c_+^2-c_-^2)]^2.
\end{eqnarray}

By combining Eqs. \eqref{eq-theo28} and \eqref{eq-theo29}, one can
note that both $Q$ and $H$ reach their maxima when $\phi=\pi/2$, for
which
\begin{eqnarray}\label{eq-theo2x}
 &&Q_{\rm max}=c_+^2+c_0^2-\sin^2\theta(c_0^2-c_-^2), \nonumber\\
 &&H_{\rm max}=[c_+^2-c_0^2+\sin^2\theta(c_0^2-c_-^2)]^2.
\end{eqnarray}
Therefore, we have $\max_{\hat{e}}h(\hat{e})=2c_+^2$, and thus
\begin{eqnarray}\label{eq-theo2y}
 N_1(\rho_{AB})=c_+.
\end{eqnarray}
This completes our proof. \hfill{$\blacksquare$}

We would like to point out here that for the two-qubit $\rho_{AB}$
with nondegenerate $\rho_A$, the calculation of the trace MIN can also
be performed in a similar manner as that for the degenerate case.
But now the expression for $h(\hat{e})$ in Eq. \eqref{eq-theo27}
is more complicated (see Eq. [31] in Ref. \cite{Ciccarello} for more
detail), and therefore we adopted the procedure listed above as the
optimal $\{\tilde{\Pi}_{1,2}^A\}$ can be written explicitly for this case.

\begin{figure}
\centering
\resizebox{0.48\textwidth}{!}{%
\includegraphics{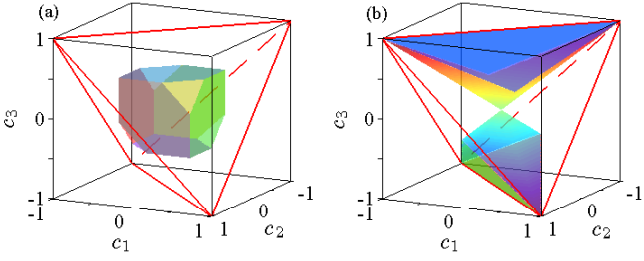}}
\caption{(Color online) Surfaces of constant trace MIN
         $N_1(\rho^{\rm BD})= 0.45$ (a), and  valid
         $(c_1,c_2,c_3)$ (the color shaded regions) for which
         $N_1(\rho^{\rm BD})$ is not destroyed by the phase
         flip noise (b).}
         \label{fig:1} 
\end{figure}

In Fig. \ref{fig:1}(a), we presented an exemplified plot of the
level surfaces of $N_1(\rho^{\rm BD})=0.45$ for the Bell-diagonal
states $\rho^{\rm BD}$ [i.e., $\vec{x}=0$ and $\vec{y}=0$ in Eq.
\eqref{eq7}]. As physical $(c_1,c_2,c_3)$ belongs to a tetrahedron
$\mathcal{T}$ (see Fig. \ref{fig:1}), and $N_1(\rho^{\rm
BD})=\max\{|c_1|,|c_2|, |c_3|\}$, the surfaces of constant trace MIN
correspond to the cross sections of the six surfaces of a cube
$\mathcal{C}$ of side length $N_1(\rho^{\rm BD})$ with
$\mathcal{T}$. When $N_1(\rho^{\rm BD})\leq 1/3$, the surfaces of
$\mathcal{C}$ are also the surfaces of constant trace MIN, while for
$N_1(\rho^{\rm BD})> 1/3$, partial of them are cut by the four
surfaces of $\mathcal{T}$.

By denoting $c_+ $, $c_0$, and $c_- $ the maximum, intermediate, and
minimum values of $\{|c_1|,|c_2|,|c_3|\}$, respectively, one can
derive a relation between $N_1(\rho^{\rm BD})$ and $N_2(\rho^{\rm
BD})=(c_+^2+c_0^2)/4$ for the Bell-diagonal states $\rho^{\rm BD}$
as
\begin{eqnarray}\label{eq16}
 N_1(\rho^{\rm BD})= \sqrt{4 N_2(\rho^{\rm BD})-c_0^2}.
\end{eqnarray}

This implies that the two different MIN measures may impose
different orderings of nonlocality, as when $c_+$ keeps unchanged,
$N_1(\rho^{\rm BD})$ also keeps unchanged, while $N_2(\rho^{\rm
BD})$ increases (decreases) with the increasing (decreasing) value
of $c_0$ in the region of $c_-\leqslant c_0\leqslant c_+$. Thus
there is no one-to-one correspondence between the well-defined trace
MIN and the conventional MIN in general, and we hope this simple
example may provide some intuition about the subtle issue concerning
the appropriateness of using the Hilbert-Schmidt norm as a distance
for quantifying nonlocality, just as the appropriateness of using it
for defining GQD \cite{Piani}.

Moreover, it is also worthwhile to point out that when $\rho^{\rm
BD}$ being subject to the $\$^{(i)}$ channel (with $i=1,2,3$
representing respectively, the bit flip, bit-phase flip, and phase
flip channels), we have $c_i(t)=c_i(0)$, and
$c_{j,k}(t)=c_{j,k}(0)p(t)$ ($i\neq j\neq k$), where
$p(t)=e^{-\gamma t}$ for the one-sided channel
$\$^{(i)}\otimes\mathbb{I}_2$ or $\mathbb{I}_2\otimes \$^{(i)}$, and
$p(t)=e^{-2\gamma t}$ for the two-sided channel $\$^{(i)}\otimes
\$^{(i)}$, with $\gamma$ being the decay rate. As a consequence, if
$|c_i(0)| =\max\{|c_i(0)|,|c_j(0)|,|c_k(0)|\}$ at the initial time,
we obtain $N_1[\$^{(i)}(\rho^{\rm BD})]=|c_i(0)|$ by Eq.
\eqref{eq8}, which is not destroyed by the $\$^{(i)}$ noise during
the whole time region. This is in sharp contrast to other
nonclassical correlation measures which remain constant only for a
finite time interval \cite{sudden}. This unique and novel
characteristic of the trace MIN is not only conceptually
significant, but is also appealing for potential quantum algorithms
relying on it.

Fig. \ref{fig:1}(b) plots the valid regions of $(c_1,c_2,c_3)$ for
which $N_1(\rho^{\rm BD})$ can evade the detrimental effects of the
phase flip channel. They belong to two hexahedra with vertices
$(0,0,0)$, $(\pm 1,\mp1,1)$, $(\pm 1/3,\pm1/3,1/3)$, and $(0,0,0)$,
$(\pm 1,\pm 1,-1)$, $(\pm 1/3,\mp 1/3,-1/3)$, respectively. The
results for the bit (bit-phase) flip channel is similar, with $c_3$
replacing $c_1$ ($c_2$).

So far we have obtained analytical formulas of the trace MIN for any
$2\times n$ dimensional pure state and a general two-qubit state,
and discussed several interesting implications of them. We now turn
to consider two high-dimensional states with symmetry. The
analytical expressions of some quantum correlation measures (see
Ref. \cite{Vedral-RMP} for a review) for them have already been
obtained \cite{gqd1,square,ana1,ana2}.

Consider first the celebrated Werner state on $\mathbb{C}^d \otimes
\mathbb{C}^d$ \cite {Werner}, which can be written as
\begin{eqnarray}\label{eq17}
 \rho^{W}=\frac{d-x}{d^3-d}\mathbb{I}_{d^2} +\frac{dx-1}{d^3-d}
          \sum_{i,j} |ij\rangle\langle ji|,~~ x\in[-1,1],
\end{eqnarray}
which admits the local unitary invariance, i.e., $\rho^W = (U\otimes
U) \rho^W (U^\dag\otimes U^\dagger)$ for any local unitary operation
$U$, and therefore one can choose the optimal measurement basis to
be $\tilde{\Pi}_i^A=|i\rangle\langle i|$, which yields
\begin{eqnarray}\label{eq18}
 \rho^{W}-\tilde{\Pi}^A(\rho^W)=\frac{dx-1}{d^3-d}\sum_{i\neq j}
                                |ij\rangle\langle ji|.
\end{eqnarray}

As $\sum_{i\neq j} |ij\rangle\langle ji|$ constitutes a permutation
matrix (a binary matrix with exactly one entry 1 in each row and
each column and zeros elsewhere), the singular values of
$\rho^{W}-\tilde{\Pi}^A(\rho^W)$ can be evaluated directly as
$|dx-1|/(d^3-d)$ with multiplicity $d(d-1)$. Then, by the definition
\eqref{eq3} we obtain
\begin{eqnarray}\label{eq19}
 N_1(\rho^{W})=\frac{|dx-1|}{d+1},
\end{eqnarray}
therefore $N_1(\rho^W)$ vanishes only when $x=1/d$, and it implies
that for the present case, the trace MIN disappears only when
$\rho^W$ reduces to the maximally mixed one. Meanwhile, the
conventional MIN for $\rho^{W}$ had also been derived analytically
\cite{glo3}, from which we obtain
\begin{eqnarray}\label{eq-ad1}
 N_1(\rho^W)=\sqrt{d(d-1)N_2(\rho^W)}.
\end{eqnarray}
This means that both $N_1$ and $N_2$ give qualitatively the same
descriptions of nonlocality for $\rho^W$ with finite $d$. But it
should be note that their asymptotic behaviors are different because
$\lim_{d\rightarrow\infty}N_1(\rho^W)=|x|$, and
$\lim_{d\rightarrow\infty}N_2(\rho^W)=0$.

The second high-dimensional state we want to consider is the $d\times d$
dimensional isotropic state expressed as follows
\begin{eqnarray}\label{eq20}
 \rho^{I}=\frac{1-x}{d^2-1}\mathbb{I}_{d^2}+\frac{d^2 x-1}{d^2-1}
          |\Phi\rangle\langle \Phi|,~~ x\in[0,1],
\end{eqnarray}
with $|\Phi\rangle=\frac{1}{\sqrt{d}}\sum_i |ii\rangle$, and
$|i\rangle$ denotes the computational basis on $ \mathbb{C}^d$. For
this state, let $\tilde{\Pi}_k^A=|\tilde{k}\rangle\langle\tilde{k}|$
be the measurement basis that maximizes the trace norm in Eq.
\eqref{eq3}, then due to the symmetry of $|\Phi\rangle$, one can
always find local unitary operation $U$ such that $(U\otimes
U)|\Phi\rangle=\frac{1}{\sqrt{d}} \sum_k |\tilde{k}\tilde{k}
\rangle$, and the local unitary invariance of $N_1$ enables
$N_1(\rho^I)=N_1[(U\otimes U)\rho^I (U^\dag\otimes U^\dag)]$.
Therefore, by denoting $\rho^{I}_U=(U\otimes U)\rho^I (U^\dag\otimes
U^\dag)$, we obtain
\begin{eqnarray}\label{eq21}
 \rho^{I}_U-\tilde{\Pi}^A(\rho^I_U)=\frac{d^2 x-1}{d(d^2-1)}\sum_{k\neq l}
                      |\tilde{k}\tilde{k}\rangle\langle\tilde{l}\tilde{l}|,
\end{eqnarray}
the singular values of which can be evaluated analytically as $|d^2
x-1|/(d^2+d)$ with multiplicity $1$ and $|d^2 x-1|/(d^3-d)$ with
multiplicity $d-1$, and this yields
\begin{eqnarray}\label{eq22}
 N_1(\rho^{I})=\frac{2|d^2 x-1|}{d(d+1)}.
\end{eqnarray}

Here, the trace MIN $N_1(\rho^I)=0$ only when $x=1/d^2$, namely,
when $\rho^I$ comes to be maximally mixed. Moreover, the analytical
expression for the conventional MIN can be obtained from Ref.
\cite{glo3}, by combining of which we obtain
\begin{eqnarray}\label{eq-ad2}
 N_1(\rho^I)=2\sqrt{\frac{(d-1)N_2(\rho^I)}{d}}.
\end{eqnarray}
Their asymptotic values are given by $\lim_{d\rightarrow\infty}
N_1(\rho^I)=2x$, and $\lim_{d\rightarrow\infty}N_2(\rho^I)=x^2$,
respectively. It implies that the two MIN measures still give
qualitatively the same characterizations of nonlocality for the
isotropic state. Moreover, it is remarkable that for the special
case $x=1$, i.e., when $\rho^I$ reduces to the maximally entangled
state, we have $N_1(\rho^{I}) =2(d-1)/d$, which is just twice that
of the conventional MIN.

\section{Summary and discussion}\label{sec:5}
To summarize, we have introduced a well-defined measure of
nonlocality by making using of the trace norm. It can remedy the
undesirable property of the conventional MIN which can be changed
arbitrarily and reversibly by trivial local action on the subsystem.
We proved explicitly that the proposed trace MIN is nonincreasing
under the action of general CPTP quantum channels on the unmeasured
subsystem. This property has by itself a conceptual significance, as
it has already been proved that the Schatten 1-norm (trace norm) is
the only $p$-norm that can be used to give a well-defined quantum
correlation measure \cite{trace}. Here, the fascinating properties
of the trace MIN show again the ubiquitousness and intrinsic
significance of the Schatten 1-norm for defining MIN. We hope this
may shed some new light on the issue concerning the characterization
and quantification of nonlocality from a measurement perspective.

We have also presented analytical formulas of the trace MIN for any
$2\times n$ dimensional pure state, two-qubit state, as well as the
Werner states and the isotropic states on $\mathbb{C}^d \otimes
\mathbb{C}^d$ which possess high symmetry. We revealed through these
results that the trace MIN captures the nonlocal property of a
system more intrinsically than that of the conventional MIN.
Moreover, we revealed a unique and appealing characteristic of this
new proposed nonlocality measure, namely, it can evade the
detrimental effects of certain noisy channels during the whole time
region for elaborately designed initial states. This may have
potential applications in QIP for its coherence protecting property.

We remark that the entropic measure of MIN based on the von Neumann
entropy \cite{min2}, or its equivalent form based on the relative
entropy \cite{min3}, is also monotonically decreasing due to the
monotonicity of the quantum mutual information under channels on $B$
(see Ref. \cite{Piani} for a detailed proof). Moreover, it has
already been pointed out that one can remedy the MIN via the square
root of the considered density matrix \cite{square}. Here, we
mention that it is also natural to define the MIN as
\begin{eqnarray}\label{eq-ad3}
 N_B(\rho_{AB})= 2\max_{\Pi^A}\{1 -\sqrt{F[\rho_{AB},\Pi^A(\rho_{AB})]}\},
\end{eqnarray}
via the Bures distance \cite{bures}, with $\Pi^A$ being the locally
invariant measurement and $F(\rho,\sigma)=[\text{Tr}(\sqrt{\rho}
\sigma \sqrt{\rho})^{1/2}]^2$ denotes the Uhlmann fidelity. By using
the monotonicity of the Bures distance \cite{Nielsen} and after a
similar analysis as that for proving Theorem 1, one can show
directly that $N_B$ is also nonincreasing under general CPTP
channels. But its evaluation may be intractable and further
investigation is needed. Thus the measure presented in this paper
may be widely used for its concise and simple form.

The significance for this computable measure of the trace MIN in QIP
can be studied parallel as that of quantum entanglement and quantum
discord, see reviews \cite{RMP,Vedral-RMP}. Additionally, it may be
applied as a new technique to various models to study physical
phenomena such as quantum phase transitions, topologies of those
systems, similar as the applications of entanglement, see for
example \cite{Amico-RMP,Cui-NC,Fan-PRL}. Moreover, as nonlocality is
quantitatively related with Heisenberg's uncertainty principle
\cite{Oppenheim-science}, which provides the basis for the security
of quantum cryptography \cite{Berta-NP}, the obvious physical
significance of the trace MIN and its convenience of calculation is
also hoped to play a role in these related issues.

\section*{ACKNOWLEDGMENTS}
This work was supported by NSFC (11205121, 11175248), the ``973''
program (2010CB922904), NSF of Shaanxi Province (2014JM1008), and
SRP of the Education Department of Shaanxi Province (12JK0986).

\newcommand{\PRL}{\emph{Phys. Rev. Lett.} }
\newcommand{\RMP}{\emph{Rev. Mod. Phys.} }
\newcommand{\PRA}{\emph{Phys. Rev.} A }
\newcommand{\PRB}{\emph{Phys. Rev.} B }
\newcommand{\PR} {\emph{Phys. Rev.} }
\newcommand{\NJP}{\emph{New J. Phys.} }
\newcommand{\JPA}{\emph{J. Phys.} A }
\newcommand{\JPB}{\emph{J. Phys.} B }
\newcommand{\PLA}{\emph{Phys. Lett.} A }
\newcommand{\NP}{\emph{Nat. Phys.} }
\newcommand{\NC}{\emph{Nat. Commun.} }
\newcommand{\QIC}{\emph{Quantum Inf. Comput.} }
\newcommand{\QIP}{\emph{Quantum Inf. Process.} }
\newcommand{\EPJD}{\emph{Eur. Phys. J.} D }
\newcommand{\AP}{\emph{Ann. Phys.} }
\newcommand{\IJMPB}{\emph{Int. J. Mod. Phys.} B }

%

%

\end{document}